\documentstyle[12pt]{article}
\begin{document}
\begin{flushright}
nlin.SI/0004009\\
SNBNCBS-2000
\end{flushright}
\vskip .7cm
\begin{center}
{\bf { Geometrical Aspects of Integrability \\
in Nonlinear Realization Scheme}}

\vskip 2cm

{\bf R.P.Malik}
\footnote{ E-mail address: malik@boson.bose.res.in  }
\footnote{ Invited talk delivered 
in a workshop on ``Dynamical Systems:
Recent Developments'' (held 
from November 4 to November 6, 1999)
at School of Physics, 
University of Hyderabad (India).}\\
{\it S. N. Bose National Centre for Basic Sciences,} \\
{\it Block-JD, Sector-III, Salt Lake, Calcutta-700 091, India} \\

\vskip 2.5cm

{\bf ABSTRACT}               

\end{center}

We discuss the integrability properties of the Boussinesq equations
in the language of geometrical quantities defined on an appropriately
chosen coset manifold connected with the $W_{3}$ algebra of Zamolodchikov. 
We provide a geometrical interpretation to the commuting conserved quantities, 
Lax-pair formulation, zero-curvature representation, Miura maps, etc.  
in the framework of nonlinear realization method.

\baselineskip=16pt


\newpage

\noindent
Two ($ 1 + 1$) dimensional integrable 
nonlinear partial differential equations have played a notable role in the
understanding of some of the physical phenomena of nature. The KdV and
Boussinesq equations, their modified versions,
their higher order  hierarchies, etc.,  are the cardinal examples of such
a class of equations which have found applications in as diverse areas
of research as two-dimensional (2D) conformal field theories, 
($W$)-string theories, fluid mechanics,
plasma physics, 2D ($W$)-gravity theories, etc.[1-4].
The latter equation (i.e., the Boussinesq equation) can be realized on
$u(x,t)$ and $ v(x,t)$ fields as
$$
\begin{array}{lcl}
\frac{\partial u} {\partial t} = - \frac{160}{3}\; v^\prime, \qquad
\frac{\partial v} {\partial t} =  \frac{1}{10}\; 
u^{\prime\prime\prime} - \frac{24}{5}\; u \; u^\prime,
\end{array}\eqno(1)
$$
which can be combined together to yield a nonlinear partial
differential equation (NLPDE) realized on the single field $u(x,t)$ as
$$
\begin{array}{lcl}
 \frac{\partial^2  u} {\partial t^2} = - \frac{16}{3}\; 
u^{\prime\prime\prime\prime} + 256 \; u \; u^{\prime\prime}
+ 256 \;u^\prime \; u^\prime, 
\end{array}\eqno (2)
$$
where the primes (i.e., $u^\prime = \frac{\partial u} {\partial x},\;
v^\prime = \frac{\partial v}{\partial x}$) denote the partial derivatives
on the fields with respect to the space variable $x$ and it has, 
as is physically evident, the dimension of length $L$
(i.e. $ x \sim L$). Taking into account this dimension, it can be readily
seen that the evolution parameter $t$, fields $u(x,t) $ and 
$v(x,t)$ have the dimensions $ L^2, L^{-2}$ and $L^{-3}$ respectively. 
In the more sophisticated language of conformal field theory, one says that 
the naive conformal dimensions of $ t, u, v$ are $ -2, +2, +3 $ respectively.
In what follows, we shall be calling it the conformal spins as is the
practice in the realm of research activities in the 
rational conformal field theories
\footnote{ 2D conformal field theory can be written in terms of the
complex variables $z$ and $\bar z$. There are holomorphic and antiholomorphic
parts in the theoy which factorize due to conformal invariance.
The conformal spin is actually defined as ($h - \bar h$)
where $h$ and $\bar h$ are the conformal weights of a primary field.}.
We shall see below how this dimensional analyses will be useful in our 
discussions for the nonlinear realization method (connected with the
group realizations on homogeneous spaces).

One of the key methods used in the discussions of the spontaneously broken
gauge theories is the coset space construction
$(G/H)$ where the total Lagrangian density of the theory is
found to be invariant under the group $G$ and the vacuum of the theory is 
found to be invariant under the subgroup $H$. This coset space 
construction turns out to be useful in the determination of
the number of Goldstone  bosons, massless gauge fields, massive gauge bosons, 
etc., in the language of group theory. 
To understand the geometry behind a symmetry group $G$, however, 
the key concept is to consider it as a group
of transformations acting on the coset space $(G/H)$ for 
the appropriately chosen
stability subgroup $H$. This was the starting point for the nonlinear 
realization method [5] applied to the spontaneously broken chiral gauge
theories in the late 60's and early 70's
where corresponding Lie algebras were found to be {\it linear}.
In the recent past, however, this
method was exploited for the discussion of geometry behind the 2D (super) NLPDE
connected with (super) W-type algebras [6,7] which are 
{\it nonlinear} to begin with. In fact, in these works, an infinite 
dimensional {\it linear} algebra was constructed from the nonlinear (super)
$W_{3}$ algebra and all the techniques of the nonlinear realization method
were exploited. In this presentation,
we shall see how some of the key properties
of integrability of the Boussinesq equation can be understood
in the language of geometry on the coset manifold.

Before we come over to the description of the coset space construction
for the $W_{3}$ algebra of Zamolodchikov, we shall dwell a bit on the basic
concepts associated with this method. A realization of a Lie group (or
corresponding Lie algebra) is a mathematical concept and it corresponds to
the validity of certain specific type of
differential equations for a group
valued function associated with the Lie group. As an example, 
it is a well known fact that the {\it linear }
realization of a compact, connected (semi)simple Lie group (or corresponding
Lie algebra) is nothing but the (matrix) representation of the Lie algebra.
To elaborate  and explain
these statements, let us begin with a set of fields $\psi_{n}$
(where $n$ is the  multiplicity) and consider the action of the group elements
$g$ of the compact Lie group $G$ such that the field $\psi_{n}$
transforms as [8, 9]
$$
\begin{array}{lcl}
g: \; \psi_{n} \;\;\;\rightarrow \;\;\; f_{n} (\psi, g),
\end{array} \eqno(3)
$$
where $f_{n} (\psi, g)$ is an abstract form of the transformed field 
$\psi_{n}$. Now one can impose all the four properties of a group on
transformation (3). For instance, the action of the identity ${\bf 1}$,
the existence of the inverse $( g^{-1} g = g g^{-1} = {\bf 1}$) 
and the closure property can be explicitly explained 
 in the language of transformations as 
$$
\begin{array}{lcl}
{\bf 1}: \; \psi_{n} \;\;\;\rightarrow \;\;\; f_{n} (\psi, {\bf 1})
= \psi_{n}, 
\end{array} \eqno(4)
$$
$$
\begin{array}{lcl}
 g \cdot g^{-1}: \; \psi_{n} \rightarrow f_{n} (\psi, g^{-1}) \rightarrow
f_{n} ( f(\psi, g^{-1}); g) \equiv \psi_{n}, 
\end{array}\eqno(5)
$$
$$
\begin{array}{lcl}
g^{-1} \cdot g:\; \psi_{n} \rightarrow f_{n} (\psi, g) \rightarrow 
f_{n} ( f(\psi, g); g^{-1}) \equiv \psi_{n},
\end{array}\eqno(6)
$$
$$
\begin{array}{lcl} 
g_{1} \cdot g_{2}: \psi_{n} \rightarrow f_{n} (\psi, g_{2}) \rightarrow
f_{n} ( f(\psi, g_{2}); g_{1}) \equiv f_{n} (\psi, g_{1} \cdot g_{2}),
\end{array} \eqno(7)
$$
where the last equation is just the closure property under 
binary operation ($\cdot$).
Here the binary operation is nothing but the transformation (3).
One can exploit this property to establish the associativity:
$$
\begin{array}{lcl}
f_{n} \Bigl ( \psi, g_{1} \cdot (g_{2} \cdot g_{3}) \Bigr ) =
f_{n} \Bigl ( \psi, (g_{1} \cdot g_{2}) \cdot g_{3} \Bigr ).
\end{array} \eqno(8)
$$
So far, the group element $g$ was treated as an abstract object, 
now we can write the 
explicit form of it in terms of a set of infinitesimal transformation
parameters $\epsilon_{\alpha}$ as
$$
\begin{array}{lcl}
g = {\bf 1} + i \epsilon_{\alpha} \Gamma_{\alpha},
\end{array} \eqno(9)
$$
where $\Gamma_{\alpha}$ are the set of generators for the above 
transformations which obey a commutation relationship for the given
Lie algebra as
$$
\begin{array}{lcl}
[ \Gamma_{\alpha}, \Gamma_{\beta} ] = i C_{\alpha\beta \gamma}
\Gamma_{\gamma}.
\end{array} \eqno(10)
$$
Here $C_{\alpha\beta\gamma}$ are the structure constants of the algebra.
The explicit form of the transformed field (with $ f_{n} (\psi, {\bf 1})
= \psi_{n},$ and eqn. (9)) is
$$
\begin{array}{lcl}
f_{n} (\psi, g) &=& f_{n} (\psi, {\bf 1} 
+ i \epsilon_{\alpha} \Gamma_{\alpha}) \nonumber\\
&\equiv & \psi_{n} + i \epsilon_{\alpha} f_{n \alpha} (\psi) + O (\epsilon^2),
\end{array} \eqno(11)
$$
where $f_{n \alpha} (\psi)$ is a group valued function and it retains the
information about the multiplicity as well as the group properties. For a 
given Lie algebra, the group valued 
functions $f_{n \alpha}$ can {\it not} be chosen in a perfectly
arbitrary way. Rather, they  obey certain specific kind of differential
equation for the given Lie algebra. For instance, using the closure 
property of equation (7), it can be seen that
$$
\begin{array}{lcl}
f_{n} (\psi, g_{1}^{-1} \cdot  g_{2} \cdot g_{1})
&=& f_{n} (f (\psi, g_{2}\cdot g_{1}); g_{1}^{-1}) \nonumber\\
& \equiv & f_{n} \bigl ( f \{ f(\psi, g_{1}) ;g_{2}\}; g_{1}^{-1}
\bigr ).
\end{array} \eqno(12)
$$
With the following inputs from the group properties (see, e.g., eqns. 4--8)
$$
\begin{array}{lcl}
f_{n} (\psi_{m}, {\bf 1}) = \delta_{nm} \psi_{m}, \quad
f_{n} ( f (\psi, g); g^{-1} ) = \psi_{n},
\end{array}\eqno(13a)
$$
$$
\begin{array}{lcl} 
f_{n} ( f (\psi, g ) = \psi_{n}  + i \epsilon_{\alpha}\;
(\frac{\partial f_{n} (\psi)} {\partial \psi_{m}}) \; f_{m \alpha}
+ O (\epsilon^2),
\end{array} \eqno(13b)
$$
we obtain the following differential equation satisfied by $f_{n \alpha}$
$$
\begin{array}{lcl}
f_{n \beta}\; C_{\alpha\gamma\beta}
= - i\; \frac{ \partial f_{n \alpha}} {\partial \psi_{m}}\; f_{m \gamma}
+ i \;\frac{\partial f_{n\gamma}} {\partial \psi_{m}}\; f_{m \alpha}.
\end{array} \eqno(14)
$$
All set of functions $f_{n \alpha} (\psi)$ that satisfy the above 
differential equation is said to provide a {\it realization} of the given
Lie algebra. For instance, it can be seen that the following
{\it linear} choice of $f_{n \alpha} (\psi)$
$$
\begin{array}{lcl}
f_{n \alpha} (\psi) = (\tau_{\alpha})_{nm}\; \psi_{m},
\end{array} \eqno(15)
$$
leads to 
$$
\begin{array}{lcl}
[ \tau_{\alpha}, \tau_{\beta} ] = i C_{\alpha\beta\gamma}\; \tau_{\gamma},
\end{array} \eqno(16)
$$
which is nothing but the matrix representation of the Lie algebra in (10). For
the spontaneously broken (chiral) gauge theories, the generators 
$\Gamma_{\alpha}$ of the full group $G$ have two parts. Some of the generators
$T_{i}$ ($ i : $ dimensionality of $H$) belong to the subgroup  
$H$ and the rest of
the generators $X_{a}$ belong to the coset space $G/H$. Thus, one can
parametrize the coset space by coset fields and the generators $X_{a}$. It
was shown in Refs. [5] that one can parametrize the coset space in
terms of {\it exponentials} and it {\it also}  provides  a {\it realization} 
for the factorized  Lie algebra under consideration. Due to the presence of
exponentials, this special realization is known as the {\it nonlinear
realization}. It can be seen that the {\it linear} term of this realization
(exponentials) is nothing but the representaion (cf. eqn. (16))
of the Lie algebra {\it if the stability subalgebra contains only identity
element of the Lie algebra}. Thus, we  see that the linear realization is a
special case of the nonlinear realization.

The first thing we note, before the application of nonlinear realization
method to the classical $W_{3}$ algebra with central extension (parametrized by
the central charge $c$) 
$$
\begin{array}{lcl}
&& [ L_{n}, L_{m} ] = (n -m) L_{n+m} + \frac{c}{12}\; (n^3 - n) 
\;\delta_{n+m, 0},\;
\;\;(- \infty \leq n, m \leq + \infty)
\nonumber\\
&& [ L_{n}, W_{m} ] = (2n -m) \;W_{n+m}, \nonumber\\
&& [ W_{n}, W_{m} ] = 16 \;(n -m) \;\Lambda_{n + m}
- \frac{c}{9} (n^3 - n) (n^2 - 4)\; \delta_{n+m, 0} \nonumber\\
&&\;\;\;\;\;\;\;\;\;\;\;\;\;\;\;\;\;\;- \frac{8}{3}\; (n - m)
(n^2 + m^2 - \frac{1}{2} n m - 4) \;L_{n+m}, 
\end{array} \eqno(17)
$$
is the fact that it is a {\it nonlinear} algebra because of the composite and 
nonlinear nature of $ \Lambda_{n} = - \frac{8}{3} \sum_{m} L_{n -m} L_{m}$.
It was thus essential to get a {\it linear} algebra out of it so that the
whole arsenal of techniques of coset space construction can be applied here.
To this goal, in Ref. [6], all the higher spin composite generators were
treated as independent generators. Invoking this idea 
immediately entails upon the $W_{3}$ algebra
to become an infinite dimensional linear algebra $W_{3}^{\infty}$ as 
$$
\begin{array}{lcl}
W_{3}^{\infty} = \{ L_{n}, W_{n}, \Lambda_{n}, \Phi_{n}
.........J_{n}^{h}...........\},
\end{array} \eqno(18)
$$
where $\Phi_{n}
\equiv (WT)_{n} $ is the conformal spin-5 composite generator and $J_{n}^{h}$ is
a generic composite generator with conformal spin-$h$. It can be readily seen
by taking a single contraction  of the OPE's
that all the higher conformal spin ($ \geq 4$) composite
generators form a closed algebra among themselves and they form an
ideal. One of the key subalgebras of this infinite dimensional algebra is the
one in which the Laurent indices of the generators with conformal spin-$h$
(i.e., $ J(z) = {\displaystyle \sum}_{n} J_{n}^{h}\; z^{-n-h} $) 
vary from $- (h -1)$ to $\infty$. For instance, for the conformal spin-2,
we have the generators: $ L_{-1}, L_{0}, L_{+1}, L_{+2}.......$ and
for the conformal spin-3, we have $W_{-2}, W_{-1}, W_{0}, W_{1}.........$
and so on and so forth. The stability subalgebra  
${\cal H}$ of our interest, 
from this truncated version of $W_{3}^{\infty}$, is
$$
\begin{array}{lcl}
{\cal H} = \{ W_{-1} + 2 L_{-1}, 
W_{0}, W_{1}, W_{2}, L_{1}, L_{0}, \Lambda_{n}
( n \geq - 3), J_{n}^{h} ( h\geq 5, n \geq - h +1) \},
\end{array} \eqno(19)
$$
where it can be easily seen that $W_{-2}$ 
is not present and  $L_{-1}$ appears in a particular linear combination
with $W_{-1}$ for the closure of the algebra. Thus, $W_{-2}$ and $L_{-1}$ 
can be taken into the coset space. It will be also noticed that all the higher
order conformal spin composite generators have been taken into the 
stability subalgebra as
they form an ideal. Now the element $g$ in
the coset space can be parametrized as
$$
\begin{array}{lcl}
g \in \frac{{\cal G}}{{\cal H}} = e^{t W_{-2}} e^{x L_{-1}} 
e^{\psi_{3}L_{3}}
\Bigl ( {\displaystyle \Pi}_{n = 4} 
e^{\psi_{n} L_{n}} e^{\xi_{n} W_{n}} \Bigr )
e^{u L_{2}} e^{v W_{3}}. 
\end{array} \eqno(20)
$$
It is an interesting point 
to note that out of all the generators in the coset space,
only two generators (i.e., $L_{-1}, W_{-2}$) commute with each other. They
have the dimensions of length as : $ W_{-2} \sim L^{-2}, L_{-1} \sim L^{-1}$.
To make the exponentials in equation (15)
dimensionless, it is clear that $t$ and $x$ must
have dimensions of length as: $ t \sim L^2, \;x \sim L$. It is 
illuminating to see that these are exactly the dimensions of $t$ and $x$ for
the Boussinesq equations which were discussed after eqn.(2). 
The commutativity of the generators $W_{-2}, L_{-1}$
ensures that the $t$ and $x$ directions are linearly independent on the
coset manifold and, therefore, they can be treated as coordinates. This feature
should be contrasted with other generators and associated 
tower of coset (Goldstone)
fields $ u, v, \psi_{3}, \xi_{4}, \psi_{4}, \xi_{5}..........$ which cannot
be treated as coordinates. Now any point on the manifold can be parametrized by
the coordinates $x$ and $t$ and all the fields can be treated as functions
of these coordinates.

The most important geometrical quantity in the framework of nonlinear
realization method is the one-differential Cartan form $\Omega = g^{-1}
d g $ (where $g \in \frac{{\cal G}}{{\cal H}}$)
in terms of which the curvature tensor, torsion, complex structure, etc.
of the coset manifold can be determined. Due to its very structure, it obeys
the following Maurer-Cartan equation
$$
\begin{array}{lcl}
d^{ext}\; \Omega + \Omega \wedge \Omega = 0,
\end{array} \eqno(21)
$$
which  is nothing but {\it the zero-curvature representation for the non-Abelain
gauge theory if we choose the 1-differential Cartan form in terms of the gauge
connections $ (A_{\mu} = A_{\mu}^{a} T^{a}) $ as: $\Omega 
= A_{\mu} dx^{\mu}$}. Here
$T^{a}$ are the generators of the Lie algebra under consideration. Now, it
can be seen that the non-Abelain curvature tensor, emerging from (21), is
zero, namely;
$$
\begin{array}{lcl}
&&F_{\mu\nu} = [ D_{\mu}, D_{\nu} ] = \partial_{\mu} A_{\nu} 
- \partial_{\nu} A_{\mu}
+ [ A_{\mu}, A_{\nu} ] = 0, \nonumber\\
&& D_{\mu} = \partial_{\mu} + A_{\mu}, \qquad \;\; A_{\mu} = A_{\mu}^{a}
T^{a}\; dx^\mu.
\end{array} \eqno(22)
$$
In fact, in the language of gauge theory, the choice $ \Omega = g^{-1} d g$ is
eaxctly like the {\it pure gauge} choice. Thus, the zero curvature 
representation is bound to be satisfied. For the truncated version of the
algebra $W_{3}^{\infty}$, we obtain
$$
\begin{array}{lcl}
\Omega = g^{-1} d g = {\displaystyle \sum}_{n = -1}^{\infty} \omega_{n} L_{n}
+ {\displaystyle \sum}_{n = -2}^{\infty} \theta_{n} W_{n} + 
\mbox{ higher spin contributions}.
\end{array} \eqno(23)
$$
As higher spin composite generators form an ideal, it is essential to know
only some of the lower order forms to obtain the dynamical equations of
motion if we exploit the ideas of Inverse Higgs-Covariant Reduction 
(IH-CR) procedure [10,11]. These lower order forms are
$$
\begin{array}{lcl}
\omega_{-1} &=& dx, \quad \omega_{0} = 0, \quad \omega_{1} = - 3 u dx
+ 160 v dt, \quad \omega_{2} = d u - 4 \psi_{3} dx + 320 \xi_{4} dt,\nonumber\\
\omega_{3} &=& d \psi_{3} + (\frac{3}{2} u^2 - 5 \psi_{4})\; dx
+ (560 \xi_{5} - 240 u v) \; dt, \nonumber\\
\omega_{4} &=& d \psi_{4} - 6 \psi_{5} dx + (896 \xi_{6} - 192 v \psi_{3}
- 768 u \xi_{4}) \;dt, \nonumber\\
&&.........................................\nonumber\\
\theta_{-2} &=& dt, \quad \theta_{-1} = 0, \quad \theta_{0} = - 6 u dt,
\quad \theta_{1} = - 8 \psi_{3} dt, \nonumber\\
\theta_{2} &=& - 5 v \;dx + (12 u^2 - 10 \psi_{4})\; dt, \qquad
\theta_{3} = dv - 6 \xi_{4} dx + (24 u \psi_{3} - 12 \psi_{5})\; dt
\nonumber\\
&&...........................................
\end{array} \eqno(24)
$$
According to the IH-CR procedure, one can set equal to zero all the components 
of the forms connected to the generators in the coset
space. In fact, these forms transform homogeneously under the left action of
the truncated version of $W_{3}^{\infty}$ symmetry and setting them equal to
zero does not spoil the symmetry of the group. These constraints
are just like the ``gauge-fixing'' conditions on the gauge-connections 
in the language of gauge theory. To make this statement more transparent,
it can be seen that the following constraints:
$$
\begin{array}{lcl}
\omega_{n} = 0, \quad \forall \quad n \geq 2, \qquad \theta_{n} = 0, 
\quad \forall \quad n \geq 3,
\end{array} \eqno(25)
$$
lead to the following kinematical and dynamical equations
$$
\begin{array}{lcl}
\psi_{3} &=& \frac{1}{4} \; u^\prime, \qquad
\psi_{4} = \frac{1}{5} \psi_{3}^\prime + \frac{3}{10}\; u^2, 
\qquad \psi_{5} = \frac{1}{6}\; \psi_{4}^\prime, \qquad 
\xi_{4} = \frac{1}{6}\; v^\prime, \nonumber\\
\dot u &=& - \frac{160}{3} v^\prime, \qquad \dot v = \frac{1}{10}\; 
u^{\prime\prime\prime} - \frac{24}{5} u \; u^\prime, \quad
\dot u = \frac{\partial u}{\partial t}, \quad \dot v = \frac{\partial v}
{\partial t}.
\end{array} \eqno(26)
$$
Thus, we see that the Boussinesq equation of eqn. (1) emerges here by the
IH-CR procedure applied on the coset manifold
as all the tower of fields can be 
expressed in terms of the essential fields $u$ and $v$ and
the derivatives on them . In the language of geometrical
properties on the coset manifold, 
{\it it can be seen that the Boussinesq equations
are nothing but the embedding conditions on a two dimensional $(x, t)$ geodesic
surface (parametrized by the basic fields $u (x, t)$ and $v (x, t)$ and the
derivatives on them) when one singles out this hypersurface from the
infinite dimensional coset manifold}.

Mathematically, the Boussinesq equation can be understood in the language of
group motion when infinite dimensional 
algebra of the infinite dimensional coset manifold reduces  covariantly to
the `covariant reduced algebra' generated by the elements of the $sl(3, R)$
algebra. In other words, the original Cartan form now reduces to a reduced
Cartan form (due to IH-CR procedure) as 
$$
\begin{array}{lcl}
\Omega = g^{-1} d g \rightarrow \Omega_{red} = g^{-1}_{red}\; d \;g_{red}
= {\displaystyle \sum}_{n = -2}^{n =2} \theta_{n} W_{n}
+ {\displaystyle \sum}_{n = -1}^{n = +1} \omega_{n} L_{n},
\end{array} \eqno(27)
$$
which satisfies the Maurer-Cartan equation: $ d^{ext} \Omega_{red}
+ \Omega_{red} \wedge \Omega_{red} = 0$. The explicit form  of this
reduced Cartan form is
$$
\begin{array}{lcl}
&&\Omega_{red} = A_{x}\; dx + A_{t}\; dt, \nonumber\\
&& A_{x} = L_{-1} - 3 u L_{1} - 5 v W_{2}, \nonumber\\
&& A_{t} = 160 v L_{1} - 6u W_{0} + W_{-2} - 8 \psi_{3} W_{1}
+ (12 u^2 - 10 \psi_{4}) W_{2}.
\end{array} \eqno(28)
$$
It can be now readily seen that the following zero-curvature condition
$$
\begin{array}{lcl}
F_{tx} = [ \partial_{t} + A_{t}, \partial_{x} + A_{x} ] = 0,
\end{array} \eqno(29)
$$
leads to the derivation of the Boussinesq equations. It will be noticed that
$A_{x}$ and $A_{t}$ are nothing but the $sl(3, R)$ valued 
{\it Drinfeld-Sokolov type Lax-pairs} in eqn. (28).
{\it In the language of geometry, these Lax-pairs can be understood 
as the projections of the reduced Cartan form along $x$ and $t$
directions of the coset manifold} (cf. eqn. (28)).
It is straightforward to notice that if the generators $ W_{2}, W_{1}, L_{1}$
are taken out from the stability subalgebra in (19), still the algebra will
be closed. Thus, we obtain a new subalgebra ${\cal H}_{1}$  from
${\cal H}$ as
$$
\begin{array}{lcl}
{\cal H}_{1} = \{ W_{-1} + 2 L_{-1}, 
W_{0}, L_{0}, \Lambda_{n}
( n \geq - 3), J_{n}^{h} ( h\geq 5, n \geq - h +1) \}.
\end{array} \eqno(30)
$$
In this case, the coset space can be parametrized as
$$
\begin{array}{lcl}
g_{1} \in \frac{{\cal G}}{{\cal H}_{1}} = e^{t W_{-2}} e^{x L_{-1}} 
e^{\psi_{3}L_{3}}
\Bigl ( {\displaystyle \Pi}_{n = 4} 
e^{\psi_{n} L_{n}} e^{\xi_{n} W_{n}} \Bigr )
e^{u L_{2}} e^{v W_{3}} e^{u_{1}L_{1}} e^{v_{1} W_{1}} e^{v_{2} W_{2}}. 
\end{array} \eqno(31)
$$
Furthermore, one can take out $W_{0}, L_{0}$ from the stability subalgebra
(30) and still algebra will be closed with only one $U(1)$ basic
generator ($ W_{-1} + 2 L_{-1}$). Now the coset element is

$$
\begin{array}{lcl}
g_{2} \in \frac{{\cal G}}{{\cal H}_{2}} = 
g_{1}\; e^{u_{0} L_{0}} e^{v_{0} W_{0}}, 
\end{array} \eqno(32)
$$
where ${\cal H}_{2}$ is given by
$$
\begin{array}{lcl}
{\cal H}_{2} = \{ W_{-1} + 2 L_{-1}, 
 \Lambda_{n}
( n \geq - 3), J_{n}^{h} ( h\geq 5, n \geq - h +1) \}.
\end{array} \eqno(33)
$$
In both the cases of $g_{1}$ and $g_{2}$, one can define a one-differential
Cartan form $\Omega_{1} = g_{1}^{-1}\; d\; g_{1}$ and $\Omega_{2}
= g_{2}^{-1}\; d\; g_{2}$ and apply the IH-CR procedure [10, 11] on it.
This leads to a covariant relationship between essential fields $u_{1}, v_{1}$
of the coset manifold (31) and $u, v$ fields of coset the manifold (20).
Similarly, one gets a relationship between essential fields $u_{0}, v_{0}$
of coset manifold (32) and $u_{1}, v_{1}$ fields of (30). These are
nothing but the so-called Miura maps. The dynamical equations on 
$u_{0}, v_{0}$ and $u_{1}, v_{1}$ fields can also be obtained due to
appropriate application of CR procedure on coset manifolds (30) and (32)
as we obtained for the essential fields $u, v$ in eqn. (26). Thus, we see
that {\it the Miura maps are nothing but the kinematical relationships
among essential fields when one goes covariantly from one coset manifold to
another one.}

It is obvious that the kinematical and dynamical relationships
can be obtained from the nonlinear realization method by application 
of the IH-CR procedure [10, 11]. For the first time, however, this procedure
was extended one step further to derive commuting conserved quantities
for the Boussinesq equations in Ref. [12]. We shall briefly dwell a bit on it.
For the derivation of the commuting conserved quantities, one has to
compute more higher order forms than the ones required for the derivation
of the dynamical equations. Some of these forms are
$$
\begin{array}{lcl}
\omega_{5} &=& d \psi_{5} + u d\psi_{3}
+ \bigl (\frac{1}{2} u^3 - 5 u \psi_{4} + 2 \psi_{3}^2  - 40 v^2 
- 7 \psi_{6} \bigr ) \; dx \nonumber\\
&+& \bigl ( 192 u^2 v - 336 u \xi_{5} - 704 \psi_{3} \xi_{4} - 160 v \psi_{4}
+ 1344 \xi_{7} \bigr ) \;dt, \nonumber\\
\omega_{6} &=& d\psi_{6} + 2 u d\psi_{4} 
+ (8 \psi_{3}\psi_{4} - 12 u \psi_{5} -8 \psi_{7})\; dx\nonumber\\
&+& \bigl (1920 \xi_{8} + 768 u \xi_{6} + 768 u^2 \xi_{4} - 
640 \psi_{4} \xi_{4} - 1664 \psi_{3} \xi_{5} \bigr )\; dt,\nonumber\\
&.&.........................................\nonumber\\
&.&.........................................\nonumber\\
\theta_{4} &=& d\xi_{4} + (3 u v - 7 \xi_{5})\; dx
+ \bigl (20 \psi_{3}^2 + 20 u \psi_{4} - 14 \psi_{6} - 8 u^3 - 80 v^2
\bigr )\; dt,\nonumber\\
\theta_{5} &=& d \xi_{5} + v du + (6 u \xi_{4} - 4 v \psi_{3} - 8 \xi_{6}) dx
\nonumber\\
&+& \bigl (56 \psi_{3}\psi_{4} + 320 v \xi_{4} + 12 u \psi_{5}
- 12 u^2 \psi_{3} - 16 \psi_{7} \bigr )\; dt,\nonumber\\
\theta_{6} &=& d \xi_{6} + 3 v d \psi_{3} 
+ \bigl ( \frac{9}{2} u^2 v - 15 v \psi_{4} - 9 \xi_{7} \bigr )\; dx\nonumber\\
&+& bigl (72 \psi_{3} \psi_{5} + 1680 v \xi_{5} + 30 \psi_{4}^2
- 360 v^2 u - 18 \psi_{8} \bigr )\; dt, \nonumber\\
&.&................................\nonumber\\
&.&................................
\end{array} \eqno(34)
$$
It can be readily seen that, due to IH-CR procedure, we can set the
$dt$ projection of the form $\theta_{5}$ equal to zero. This leads to
the following equation
$$
\begin{array}{lcl}
\frac{\partial \xi_{5}} {\partial t}
= 16 \psi_{7} + 12 u^2 \psi_{3} - 12 u \psi_{5} 
- 56 \psi_{3}\psi_{4} - 320 v \xi_{4}.
\end{array} \eqno(35)
$$
Now using the kinematical relationships, it can be seen that the r.h.s.
of the above expression is a total space derivative
$$
\begin{array}{lcl}
\frac{\partial \xi_{5}} {\partial t}
= \frac{\partial} {\partial x}\;
\bigl [ 2 \psi_{6} + \frac{1}{5} u^3 - 2 u \psi_{4} - \frac{16}{3} \psi_{3}^2
\bigr ],
\end{array} \eqno(36)
$$
which ultimately leads to the following conservation law
$$
\begin{array}{lcl}
\frac{\partial (u v)} {\partial t}
= \frac{\partial} {\partial x}\;
\bigl [ - \frac{11}{5} u^3 + 2 u \psi_{4} - \frac{16}{3} \psi_{3}^2
- \frac{80}{3}\; v^2
\bigr ],
\end{array} \eqno(37)
$$
where we have used $\xi_{5} = \frac{\xi_{4}^\prime}{7} + \frac{3}{7} u v,
\psi_{6} = \frac{1}{7} \bigl ( \psi_{5}^\prime + 2 \psi_{3}^2 - u^3 - 40v^2
\bigr )$ from the kinematical relationships. Similarly other conserved
quantities can be calculated (see, e.g., Ref. [12] for details). Some
of the conserved quantities are
$$
\begin{array}{lcl}
H_{1} &=& \frac{c}{2}\; {\displaystyle \int} dx\; u (x,t), \quad
H_{2} = - \;\frac{40c}{3} {\displaystyle \int} dx\; v (x, t), \quad
H_{4} = c\; {\displaystyle \int} dx\; (uv) (x, t), \nonumber\\
H_{5} &=& - \;c\;
{\displaystyle \int} dx\; \bigl [\;
\frac{(u^\prime)^2}{20} + \frac{ 4 u^3}{5} + \frac{80 v^2}{3}
\;\bigr ], \nonumber\\
H_{7} &=&  \;c\;
{\displaystyle \int} dx\; \bigl [\;
\frac{(u^{\prime\prime})^2}{3200} + \frac{ 9 u 
(u^{\prime})^2}{400} + \frac{(v^{\prime})^2} {6} + \frac{3 u^4}{50}
+ 4 u v^2\;
\bigr ], \nonumber\\
&&................................\nonumber\\
&&.................................
\end{array} \eqno(38)
$$
Here the subscripts for the conserved quantities stand for the naive
conformal dimensions (i.e. $ H_{1} \sim L^{-1}, H_{2} \sim L^{-2}$ etc.).
The commutativity of the conserved quantities (i.e., $ \{ H_{i}, H_{j} \}
= 0, \;\; i, j = 1, 2, 4, 5, 7.....$) can be established if we exploit
the following second Hamiltonian structure associated with $u$ and $v$
fields for the classical $W_{3}$ Poisson brackets
$$
\begin{array}{lcl}
\{ u(x,t), u(y,t) \} &=& \frac{2}{c}\;
\bigl [\; \frac{1}{6} \frac{\partial^3} {\partial y^3} - 2 u(y) \frac{\partial}
{\partial y} - \frac{\partial u} {\partial y} \;\bigr ] \; \delta (x -y),
\nonumber\\
\{ u(x,t), v(x,t) \} &=& - \frac{2}{c} \;\bigl
[\; 3 v(y) \frac{\partial} {\partial y} + \frac{\partial v}{\partial y}\;
\bigr ]\; \delta (x -y), \nonumber\\
\{ v(x,t), v(y, t) \} &=&
\frac{3}{100c}\; \bigl [\; - \frac{1}{48} \frac{\partial^5}{\partial y^5}
+ \frac{5}{4} u(y) \frac{\partial^3}{\partial y^3} + \frac{15}{8}
\frac{\partial u}{\partial y} \frac{\partial^2} {\partial y^2} \nonumber\\
&+& \bigl ( \frac{9}{8} \frac{\partial^2 u}{\partial y^2} - 12 u^2 \bigr )
\frac{\partial}{\partial y} + \bigl ( \frac{1}{4} \frac{\partial^3 u}
{\partial y^3} - 12 u \frac{\partial u}{\partial y} \bigr )\; \bigr ]
\delta ( x - y).
\end{array} \eqno(39)
$$
It will be noticed that there are no conserved quantities for the
conformal dimensions 3, 6, 9, 12.....( n = 0 mod 3). This is
in agreement with the famous Lenard recursion relations for the
commuting conserved quantities for the Boussinesq equations. These
conserved quantities can be understood in terms of the generators
of the infinite dimensional algebra $W_{3}^{\infty}$. For instance,
if we take the Laurent mode decomposition for the $u$ and $v$
fields and consider the holomorphic and antiholomorphic parts
together, the contour integration in (38) will lead to the
following set of generators modulo some constant factors:
$$
\begin{array}{lcl}
\{ L_{-1}, W_{-2}, \Phi_{-4}, S_{-5}.................\}
\end{array}\eqno(40)
$$
where $ \Phi = \frac{48}{c} (TW), S = \frac{1}{c}\;\bigl
( W^2 - \frac{128}{3c} T^3 + \frac{4}{3}\;(\partial T)^2 
\bigr )$.......etc. These generators form a {\it Cartan subalgebra}
in the infinite dimensional algebra $W_{3}^{\infty}$ as they
commute among themselves. In the framework of nonlinear 
realization method, these generators correspond to the translation
generators on the infinite dimensional coset manifold. 
For instance, the first conserved quantity $H_{1}$ corresponds
to the generator $L_{-1}$ which is nothing but the space
translation (momentum) generator. The rest of the conserved 
quantities correspond to the time evolution generators
on the coset manifold.  Their
commutativity corresponds to the linear independence of the
directions of the space $x$ and all the other
``time'' evolution parameters with one another.

To summarize, we have shown that: (i) the Boussinesq equations
are the embedding conditions on a 2D geodesic hypersurface
in the infinite dimensional coset manifold, (ii) Miura maps
are the covariant kinematical relationships among the essential
fields as one goes covariantly from one coset manifold to another,
(iii) Lax-pairs are the components of projections of the
reduced Cartan forms along $x$ and $t$ directions of the 
coset manifold, (iv) Commuting conserved quantities are the
translation generators on the manifold and they form a 
Cartan subalgebra in the infinite dimensional algebra 
$W_{3}^{\infty}$, (v) Commutativity of the conserved quantities
are reflected in the linear independence of all the
evolution directions $( x, t, t^\prime.....)$ on the coset manifold.

\end{document}